\documentclass[11pt,twoside]{article}

\usepackage{asp2006}
\usepackage{epsf}
\usepackage{graphicx}
\usepackage{lscape}

\begin{document}
\title{The X-ray nature of the LINER nuclear sources}   
\author{I. M\'arquez$^1$, O. Gonz\'alez-Mart\'{\i}n$^1$, J. Masegosa$^1$, M.A. Guerrero$^1$, D. Dultzin-Hacyan$^2$}   
\affil{$^1$Instituto de Astrof\'{\i}sica de Andaluc\'{\i}a (CSIC), Granada (Spain)\\$^2$Instituto de Astronom\'{\i}a, UNAM, M\'exico D.F. (M\'exico)}    

\begin{abstract} 
The analysis of the X-ray data for a sample of 
51 LINER nuclei with available X-ray Chandra imaging is reported. 
Our aim was to investigate
the physical mechanisms which power LINER nuclear activity. 
The use of multiwavelenght information at radio, UV, 
optical HST and X-ray lead us to conclude that at least 60\% of the LINERs 
are hosting a low luminosity AGN in their nuclei. This percentage may be even higher if the Compton-thickness of some nuclei (mostly with SB-like hard X-ray morphology) is confirmed.
\end{abstract}



\section{Introduction}

Pioneering works already estimated that at least 1/3 of all the spiral
galaxies are LINERs (Heckman et al. 1980). If they represented the
faint end of the AGN luminosity distribution, their study would be
crucial for the understanding of AGN activity, in particular in our
nearby universe. But still nowadays, there is an ongoing strong debate on
the origin of the energy source in LINERs, with two main alternatives
for the ionizing source: either it is a low luminosity
AGN, or it has a thermal origin related to massive star formation
and/or from shock heating mechanisms resulting from the massive stars
evolution.  The search for a compact X-ray nucleus in LINERs is indeed
one of the most convincing evidence about their AGN nature.

\section{The sample and the data}
The catalogue by Carrillo et al. (1999) provides 476 LINERs with
information from radio to X-rays. Our sample comprises all the LINERs
in this catalogue with enough quality data (at least 25 counts in the
range 0.5-10 keV) in Chandra-ACIS archives, made public up to Novembre
2004. This sample amounts to 51 LINERs, which resulted to be
representative of the bright optical LINERs (see Gonz\'alez-Mart\'{\i}n et
al. 2006, for details).

\section{X-ray Analysis}
Attending to their hard X-ray (4.5-8.0*keV band) morphology the sample
has been grouped into two categories: (a) {\bf AGN-like Nuclei}, with
unresolved nuclear point-like source, and (b) {\bf Starburst-like Nuclei},
with no detection of a nuclear point-like source.

The spectra from 0.5 to 10 keV have been fitted.
Two models (thermal and non-thermal), and their combination, were
tested to account for the spectral emission of the 23 objects with
enough S/N, including absorption (phabs) and using solar metallicities
and $\chi$$^2$ statistics.  Only one object can be better fitted with
a pure thermal model, whereas for 7 of them a pure power law is a
better description. A combination of thermal and non-thermal
components is required in the remaining cases. The resulting median
spectral parameters are kT=0.64$\rm{\pm 0.17}$~keV and
$\Gamma$=1.89$\rm{\pm 0.45}$.

\begin{figure}
\begin{center}
\includegraphics[width=0.30\textwidth]{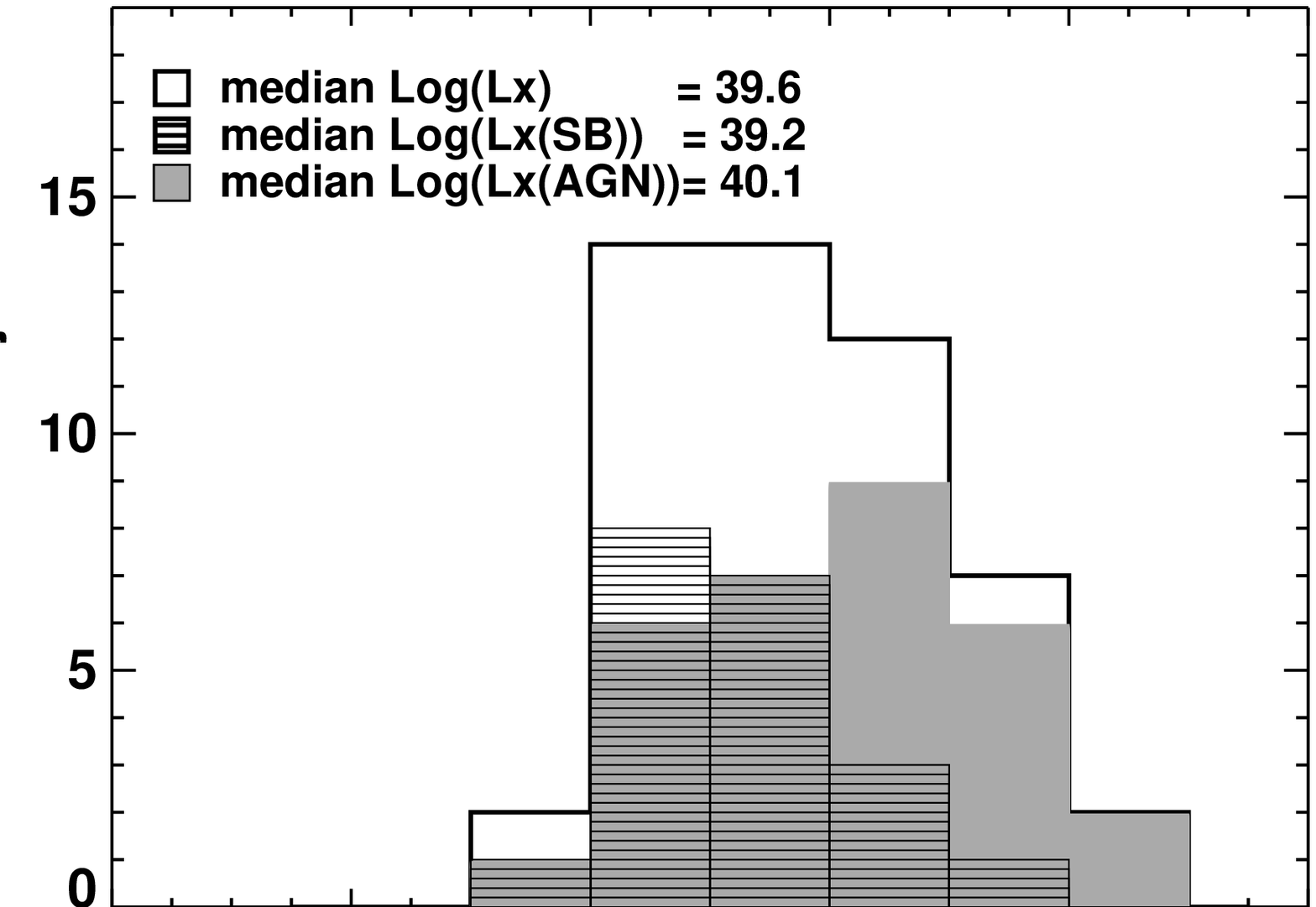}
\includegraphics[width=0.30\textwidth]{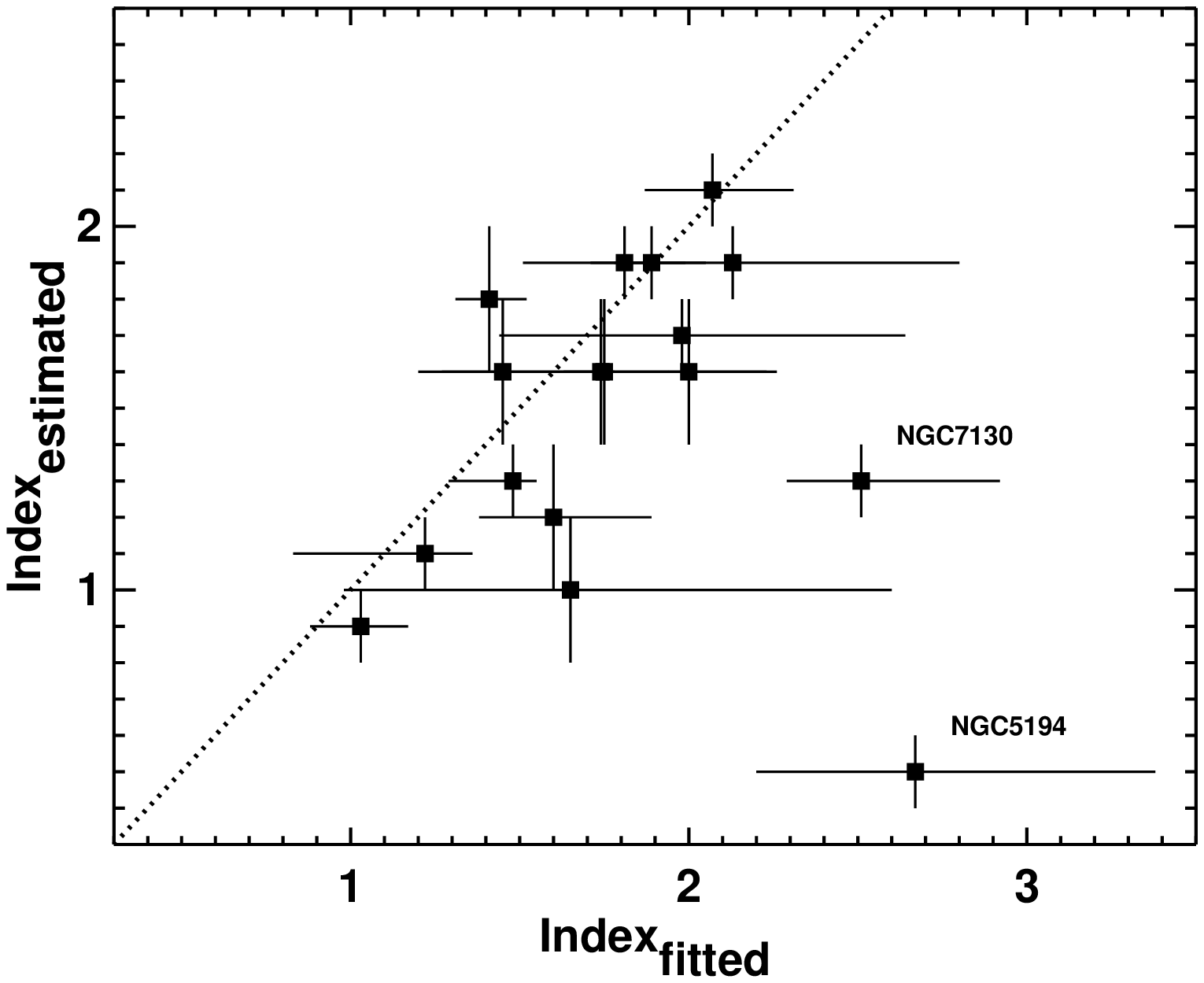}
\includegraphics[width=0.30\textwidth]{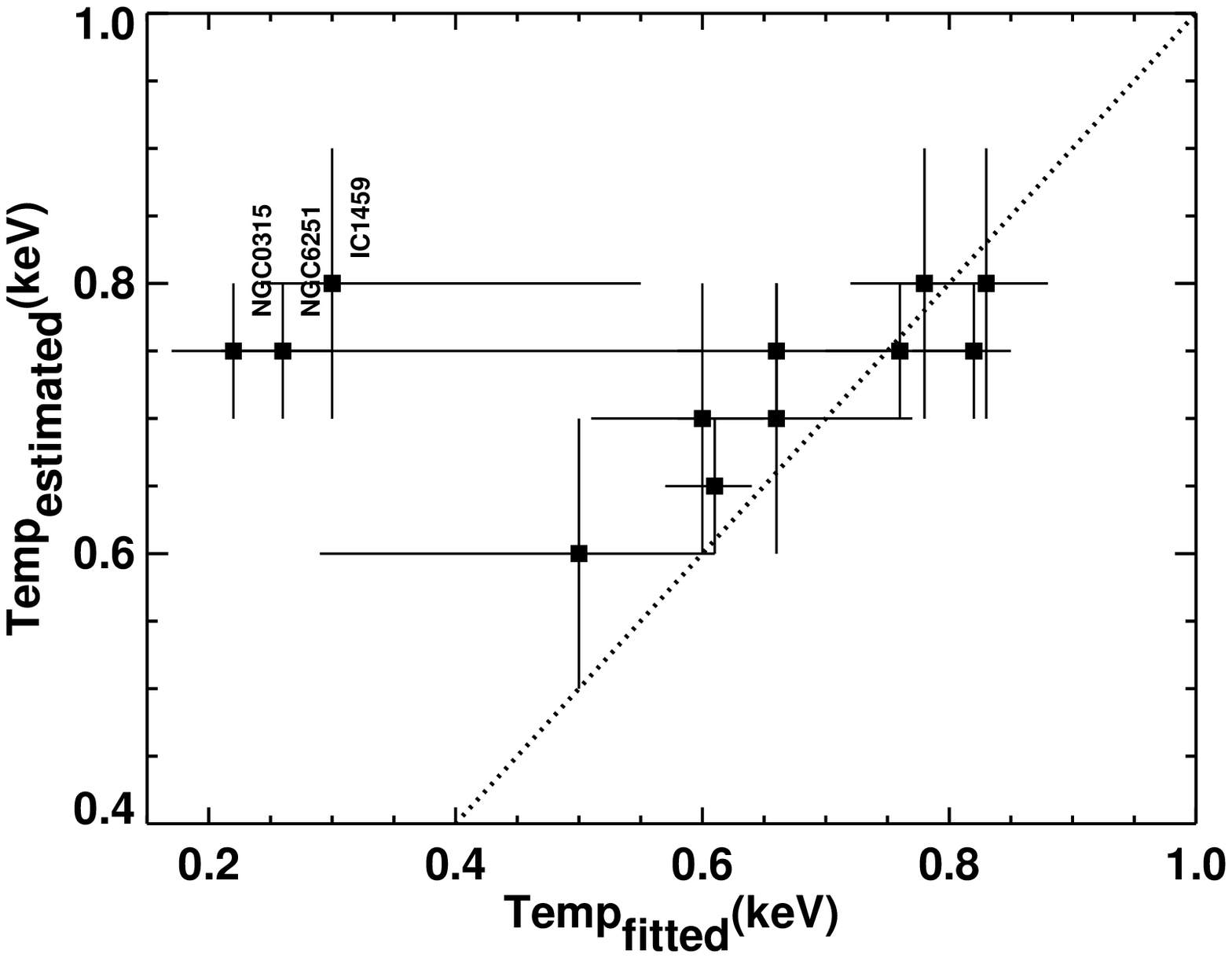}
  \caption{(Left): Luminosity histogram for the whole sample
 (empty); objects classified as AGN-like (grey) 
 and Starburst-like objects (dashed). Comparison between power-law 
 index (centre) and temperature (right) as estimated from C-C
 diagrams with those from fitted values.}\label{fig:fig1}
\end{center}
\end{figure}

Nuclear luminosities were calculated from the best-fit model for these
26 galaxies, or estimated assuming a power law of $\Gamma=1.8$ and
galactic absorption for the remaining 28 galaxies, for which no
spectral fitting is possible. Fig. \ref{fig:fig1} (left) shows that
AGN-like nuclei tend to be more luminous than SB-like nuclei.

Color-color (C-C) diagrams have been defined and optimized to estimate
$\Gamma$, kT, and n$\rm{_{H}}$ when the spectral fitting is not
possible. Fig
\ref{fig:fig1} shows the comparison between estimated and fitted
values.  Excepting three cases, the C-C diagrams provide a good T
estimation, and $\Gamma$ results to be slightly underestimated.

\section{HST imaging}
Sharp-divided images of the objects with available HST imaging (43
galaxies) are used to classify our sample LINERs into two groups: 
(a) compact nuclear sources
(35 objects); and (b) dusty nuclear regions (8). All the galaxies
classified as AGN by the X-ray imaging analysis show compact nuclei at
the resolution of HST images.

\section{Results}
From the X-ray analysis, about 60\% of the LINERs in our sample are
AGN candidates. These objects are, on average, more luminous in X-rays
(2-10 keV), but their luminosities share a large overlapping range
with SB candidates. Both the luminosities and the morphological
classes generally agree with those in previous studies (Ho et
al. 2001, Satyapal et al. 2004, 2005, Flohic et al. 2006). The
temperatures for the soft thermal component (kT$\approx$ 0.6-0.8 keV)
are similar to those in starbursts (Ott et al. 2005, Grimes et
al. 2005), and higher than those found for Seyfert 1 nuclei
(kT$\approx$ 0.25 keV, Panessa et al. 2006). The hard power-law slope
for LINERs ($<\Gamma>$=1.8) is within the range found for Seyferts
(Cappi et al. 2006 have found $<\Gamma>$=1.56 and 1.61 for Sy1 and
Sy2, respectively).

To gain insight into the emission mechanisms in these objects, 
we looked
for other evidence of the AGN nature at other wavelenghts:

{\bf HST Analysis}: All AGN candidates show compact HST nuclei
(28 galaxies). SB candidates (15) show both dusty (8) and compact (7)
HST nuclei. Among the seven SB candidates with unresolved optical
nuclei, 4 show other evidences of hosting AGN both in optical (broad
or double-peaked H$\alpha$ emission line) and radio (see below).

{\bf Radio Evidence}: An unresolved nuclear radio core and flat
continuum have been suggested as an evidence of AGN nature. Most of
the LINERs in our sample classified by Filho et al. (2000) as AGNs in
radio (13 objects) result in AGN-like class in X-rays (9).

{\bf UV variability}: For the 7 objects in common with the sample of 
17 LINER galaxies with HST/UV data by Maoz et al. (2005), 
five show time variability hinting to their AGN nature; all of them 
belong to our AGN-like class. 

{\bf Stellar populations}: For the 14 galaxies with data available
 (Cid-Fernandes et al. 2004) the contribution of young stars is always
 less than 3\%. Therefore, High Mass X-ray Binaries are not expected
 to be an important ingredient for the nuclear X-ray emission.

\section{Compton Thickness} 
\begin{figure}
\begin{center}
\includegraphics[width=0.40\textwidth]{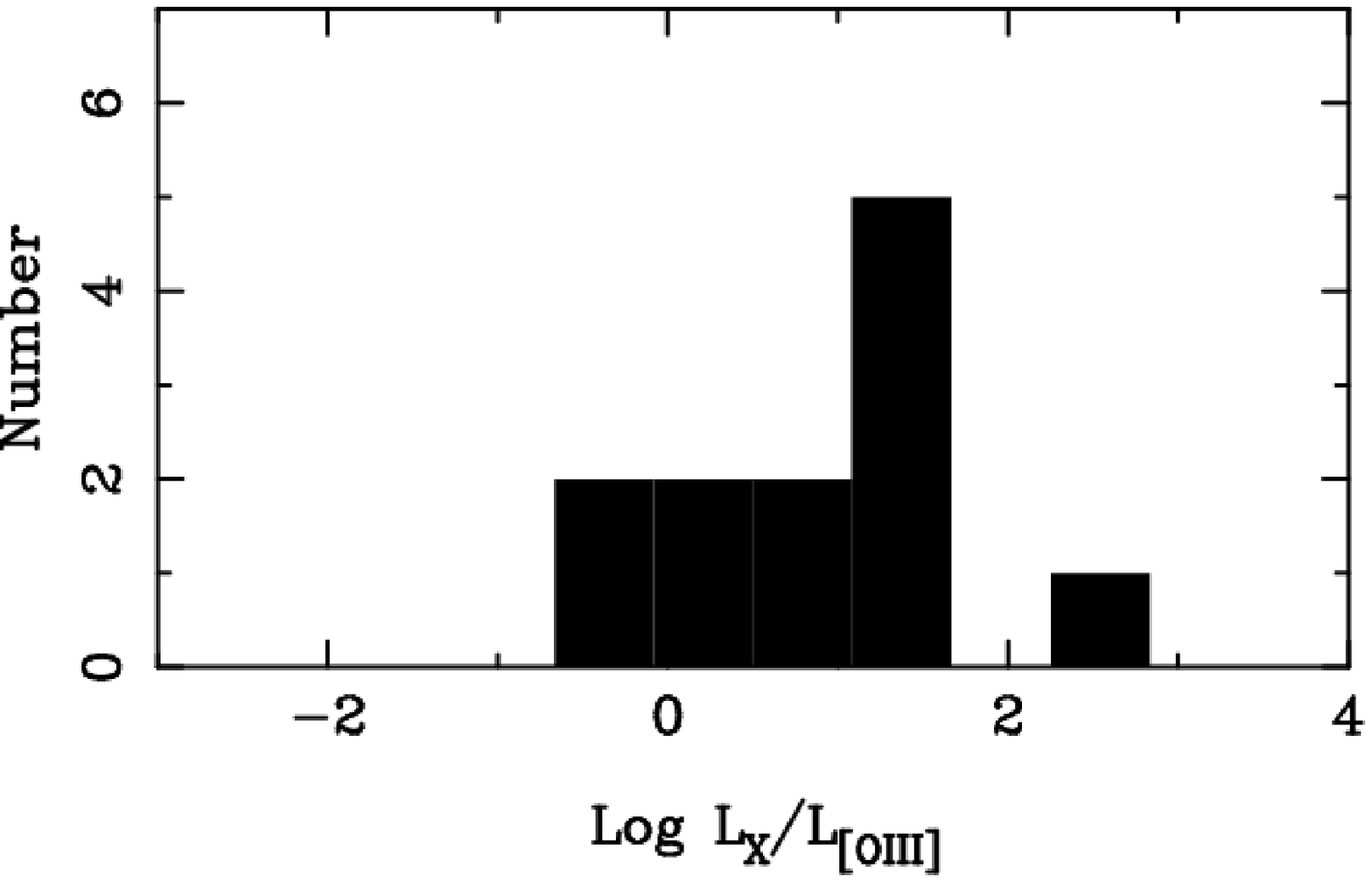}
\includegraphics[width=0.41\textwidth]{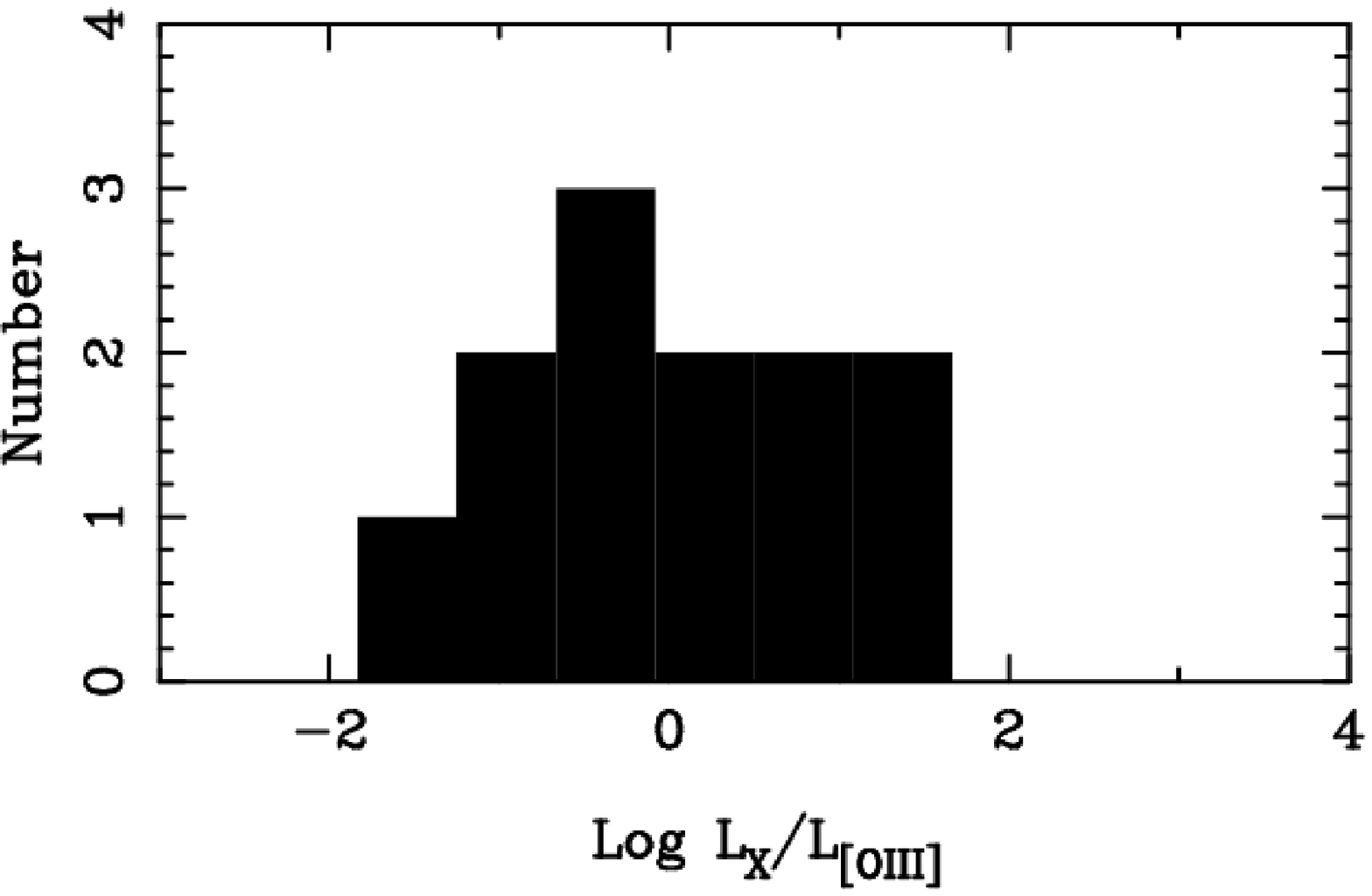}
  \caption{L$_{\rm X}$/L[OIII] ratios for AGN (left) and SB (right) candidates in our LINER sample. Compton-thick sources should have log(L$_{\rm X}$/L[OIII]) $<$ 0.}
\label{thick}
\end{center}
\end{figure}

L$_{\rm X}$/L[OIII] may be used to detect Compton-thick sources
(L$_{\rm X}$/L[OIII] $<$ 1, Maiolino et al. 1998). In Fig. \ref{thick}
this ratio is shown for the 24 LINERs in our sample with available
[OIII] luminosities, separately for AGN (left) and SB (right)
candidates. It can be seen that SB-like nuclei show lower values than
AGN-like (mean values 0.96 and 11.0, respectively).  The high
percentage of SB-like in the region occupied by Compton-thick objects
(50\%) imply that they may host a strongly obscured AGN. This issue
will be further analysed by searching for the presence of Fe lines in 
XMM spectra.

\section{Conclusions}\label{sec:concl}
From the study of the nuclear properties of a sample of galaxies with LINER 
nuclei we have obtained that:
\begin{itemize}
\item From the X-ray morphology, 60\% of the LINERs are classified as AGN-candidates. 
\item Most of the objects need both thermal and non-thermal components for the 
 spectral fitting. C-C diagrams confirm this result.
\item All the galaxies classified as AGN by the X-ray imaging show Compact 
 nuclei in the optical at the HST resolution. 
\item Nine of the objects showing  AGN nature at radio frequencies belong 
 to our AGN-like class.
\item Seven LINERs are variable at UV; all of them are AGN-like. 
\item High Mass X-ray Binaries are not expected to be an important contribution 
 to the X-ray emission.
\item A high percentage of the nuclei in our SB-like class may host strongly 
 obscured AGNs.
\end{itemize}

Although contributions from HMXBs and ULXs cannot be ruled out for
some galaxies, we concluded in Gonz\'alez-Mart\'{\i}n et al. (2006) that
60\% seems to be a lower limit for LINERs hosting an AGNs. The
analysis of the Compton Thickness hints to a
high percentage of our SB-like LINER nuclei hosting strongly obscured AGNs.

\acknowledgements 
This work was financed by DGICyT grant AYA2003-00128 and the Junta de
Andaluc\'{\i}a TIC114. OGM acknowledges financial support by the Spanish 
Ministerio de Educaci\'on y Ciencia through the grant FPI
BES-2004-5044.



\end{document}